# Two-dimensional β-phase copper iodide: a promising candidate for low-temperature thermoelectric applications

Bingquan Peng[1*], Yinshuo Li[2], and Liang Chen[2, 3*]

[1] Wenzhou Institute, University of Chinese Academy of Sciences, Wenzhou, Zhejiang 325000, China;

[2] Zhejiang Provincial Key Laboratory of Chemical Utilization of Forestry Biomass, Zhejiang A&F University, Lin'an, Zhejiang 311300, China

[3] School of Physical Science and Technology, Ningbo University, Ningbo 315211, China;

[*]Corresponding authors: E-mail: pengbq@ucas.ac.cn (B.Q.P); liangchen@zafu.edu.cn (L.C.);

## ABSTRACT

Bismuth telluride-based materials is the only commercially viable room-temperature thermoelectric material, despite its limited tellurium and poor mechanical properties. The search for materials with a high figure of merit ($zT > 1.00$) near room temperature remains a major challenge. In this work, we systematically investigate the structural stability and the thermoelectric capabilities of monolayer β-CuI and γ-CuI through the density functional theory (DFT) combined with Boltzmann transport theory. Based on the thermoelectric transport coefficients of monolayer β-CuI and γ-CuI, we predict their zT values will vary with carrier concentration and increase with temperature. Comparing the *zT* values, monolayer β-CuI demonstrates superior thermoelectric properties compared to γ-CuI. At room temperature, the optimal *zT* values of monolayer β-CuI exceed 1.50, with particularly high values of 2.98 (p-type) and 4.10 (n-type) along the Zigzag direction, demonstrating significant anisotropy. These results suggest the great potential of the monolayer β-CuI is promising candidate materials for low temperature thermoelectric applications.

# INTRODUCTION

Thermoelectric materials directly convert heat energy into electricity and vice versa, efficiently harnessing solar, geothermal, and waste heat for power generation and serving as a heat pump for cooling.[1-5] Due to its rapid response, zero emissions, and excellent recyclability, thermoelectric technology based on thermoelectric materials is increasingly recognized as a promising solution for reducing $CO_2$ and greenhouse gas emissions and providing cleaner forms of energy.[6, 7] The efficiency of thermoelectric power generation depends on the material's thermoelectric figure of merit ($zT$), which is described by the equation $zT = S^2 \sigma T / (\kappa_L + \kappa_e)$, where $S$, $\sigma$, $\kappa_L$, $\kappa_e$, and $T$ represent the Seebeck coefficient, electrical conductivity, lattice thermal conductivity, electronic thermal conductivity, and absolute temperature respectively.[4] Based on extensive research, the ideal thermoelectric material is one with high $S$ and $\sigma$ values and low $\kappa$ values.[8, 9] However, the $zT$ value varies with temperature, causing different materials to exhibit their highest $zT$ values at different operating temperatures. In recent decades, significant advancements in synthesizing high-performance thermoelectric materials have achieved $zT$ values near or above 2.00[10-14], such as $Cu_2Se$,[15] $SnSe$.[16] However, most of these materials exhibit peak $zT$ only in medium- to high-temperature ranges, with lower $zT$ values near room temperature. Recent reports indicate that PbSe-based thermoelectric materials, which are being explored as alternatives to $Bi_2Te_3$, achieve exceptional electrical transport performance of approximately 52 µW cm$^{-1}$ K$^{-2}$ for cooling applications, while also demonstrating a $zT$ value of around 0.9 at room temperature.[17] Thus, the quest for materials exhibiting a $zT > 1.00$ at or near room temperature continues to pose a significant challenge.[9, 18, 19] For a long time, p/n $Bi_2Te_3$ is the only commercially available high performance ($zT \sim 1.00$) material system for near room temperature thermoelectric materials and thermoelectric refrigeration.[20] Bismuth telluride ($Bi_2Te_3$), a prominent low-temperature thermoelectric material, faces significant challenges due to the scarcity of tellurium, which is present at only 0.001 ppm, thereby severely restricting production capacity.[21-23] Consequently, there is a pressing need to develop alternative, cost-effective, and environmentally friendly materials for low-temperature thermoelectric applications.

Under ambient conditions, CuI primarily exists in the form of γ-CuI.[24, 25] As the temperature increases, it will first transition to β-CuI, and then transition to α-CuI.[25] The layered β-CuI was originally found in a very narrow high temperature range, so it is considered a high temperature phase CuI.[26] Based on phonon spectra and AIMD dynamics, it has been theorized that β-CuI can exist stably at room temperature.[27-29] It is predicted to exhibit several unique physical properties, including low thermal conductivity,[27] low elastic modulus,[28, 30] high transparency,[31] among others.[32] Recent findings indicate that monolayer β-CuI can form heterojunctions either within a graphene-confined environment or on a substrate under ambient conditions.[33, 34] Additionally, self-supported monolayer β-CuI has been observed to remain stable under ambient conditions.[35, 36] These provide impetus for further exploration of its other properties.

In recent years, a range of two-dimensional (2D) thermoelectric materials show remarkable thermoelectric properties attributed to quantum confinement effects.[37-41] Despite the fact that γ-CuI, the first transparent conductive material discovered, was reported and studied extensively,[42-50] the thermoelectric properties of the 2D layered β-CuI remain unexplored. Low-dimensional materials such as 2D materials exhibit strong quantum confinement effects,[37, 51-54] leading to significantly enhanced thermoelectric properties.[55] Additionally, as a 2D material, monolayer β-CuI is theoretically predicted to exhibit lower lattice thermal conductivity compared to bulk γ-CuI.[27] AIMD simulations demonstrate that monolayer β-CuI retains thermodynamic stability near room temperature. Exploring the thermoelectric properties of 2D β-CuI is crucial. This study employs first-principles calculations and Boltzmann transport theory to investigate the thermoelectric characteristics of monolayer β-CuI and γ-CuI. The results show that monolayer β-CuI has excellent low temperature thermoelectric performance compared with γ-CuI. Consequently, the optimal $zT$ values for n-type (p-type) doped monolayer β-CuI along the Zigzag (ZZ) and Armchair (AC) directions is 6.35 (4.83) and 2.79 (3.03) at 500 K, respectively. Most importantly, the optimal $zT$ values of monolayer β-CuI exceed 1.50 at room temperature in all cases. These findings highlight the significant potential of monolayer β-CuI as a promising candidate material for low-temperature thermoelectric applications.

## RESULTS AND DISCUSSION

### Geometrical structures and structural stability of the monolayer β-CuI

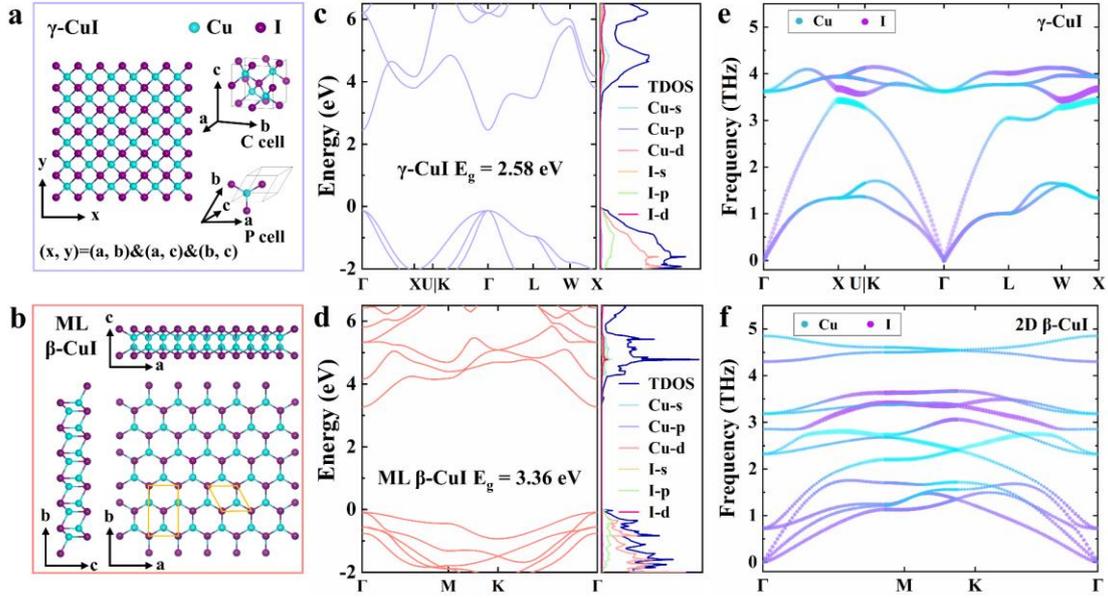

**Fig. 1** The geometric, electronic band structure with partial density of states and phonon spectra. Three crystal views of the crystal structures of (a) γ-CuI and (b) monolayer β-CuI (ML β-CuI), along with their primary and orthogonal conventional cell. Purple and cyan spheres represent I atoms and Cu atoms, respectively. The calculated band structure and total and partial density of states for (c) γ-CuI, (d) monolayer β-CuI. The phonon spectra of (e) monolayer β-CuI and (f) γ-CuI.

Fig. 1(a) presents three perspectives of the supercell of the γ-CuI conventional cell, along with the structures of its primitive cell and orthogonal conventional cell. The conventional cell of γ-CuI (space group: $F$–43m) has optimized lattice constants of 6.082 Å along the a, b, and c axes. The conventional cell of monolayer β-CuI is also orthogonal, with the a and b axes aligned along ZZ and AC directions, as depicted in Fig. 1(b). The primitive cell of monolayer β-CuI (space group: $P$–3m1) has optimized lattice constants where a = b = 4.181 Å, with a monolayer thickness of approximately

7.934 Å. The band structures and density of states (DOS) of monolayer β-CuI and bulk γ-CuI have been extensively reported in the literature.[28, 56] To better describe these band structures, we used HSE06 and obtained results consistent with previous reports,[28, 30] as depicted in Fig. 1(c) and (d). bulk γ-CuI and monolayer β-CuI are both direct bandgap semiconductors, with bandgaps of 2.58 eV and 3.36 eV, respectively. The experimentally measured bandgap of bulk γ-CuI is approximately 3.1 eV.[57-69] Despite having such a wide bandgap, bulk γ-CuI still exhibits excellent thermoelectric performance.[42, 43] The phonon dispersion spectra indicate the thermodynamic stability of γ-CuI and monolayer β-CuI, as there are no imaginary frequencies observed, depicted in Fig. 1(e) and (f), respectively. The highest frequencies of γ-CuI and monolayer β-CuI phonon vibrations are 4.13 THz and 4.85 THz, respectively, which are consistent with the reported results.[27]

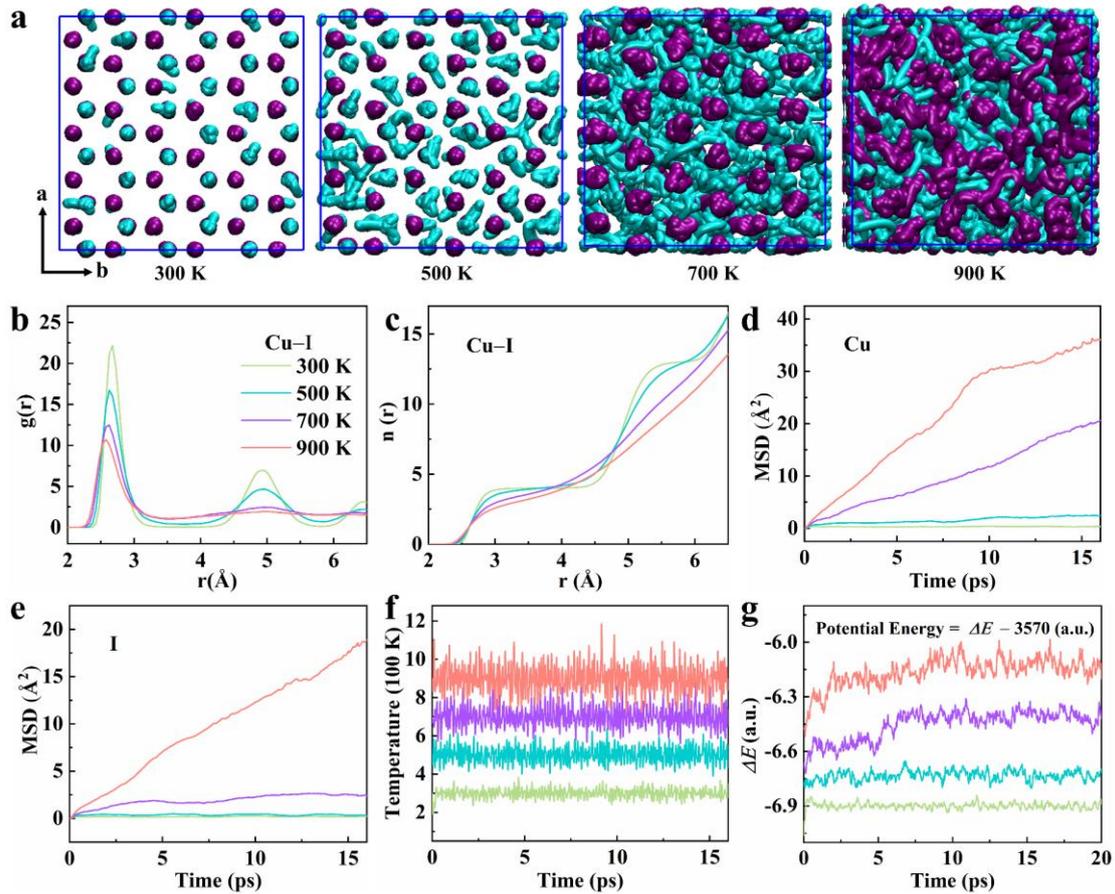

**Fig. 2** The thermal stability and phase transitions of monolayer β-CuI from AIMD simulations at different temperatures. (a) Trajectories of Cu (cyan) and I (purple) ions in a supercell from AIMD simulations at different temperatures. 2D diffusion pathway and diffusion in the ab plane. (b) Equilibrated radial distribution function (RDF) and (c)

coordination number (CN) at different temperatures. Average mean square displacements (MSDs) for (d) Cu ions and (e) I ions at different temperatures, respectively. (f) The temperatures and (g) potential energies of a monolayer β-CuI supercell are monitored during AIMD simulations at different temperatures (300 K, 500 K, 700 K, and 900 K).

It has been found that layered β-CuI is stable at room temperature, warranting further investigation into its stability and phase transitions at higher temperatures.[35, 36] Extensive AIMD simulations were conducted in the temperature range of 300 K to 900 K to investigate the stability and phase transition characteristics of monolayer β-CuI. 2D monolayer β-CuI crystal features were differentiated at different temperatures by comparing atomic trajectories in the simulations, as depicted in Fig. 2(a). The radius distribution functions (RDFs) between Cu and I were analyzed in Fig. 2(b), with an appreciable probability density observed between the first two peaks as the temperature increased, indicating the ease of Cu ions diffusion. Above 700 K, only one coordination peak was observed between Cu and I, suggesting a large degree of disorder in the Cu ions distribution and diffusing as liquid-like Cu ions. The mean square displacements (MSD) of Cu and I ions at different temperatures were analyzed. As the temperature increased, it was observed that Cu ions diffused more readily. At 700 K, Cu ions exhibited significant expansion, while I ions vibrated around their equilibrium positions, exhibiting a superionic state. Diffusion coefficients (D) were calculated from the MSDs of Cu and I ions, as shown in Fig. 2(d) and (e) (with more details in Table S1). Therefore, monolayer β-CuI exhibited a solid phase at 300 K, a superionic phase at 500 K, and liquid disordered phases at 700 and 900 K. By examining the potential energy changes in the simulation system at various temperatures, it is evident that the potential energy is stable at 300 and 500 K, while it shows significant fluctuations at 700 and 900 K compared to the lower temperatures.

**Relaxation time and electronic transport property**

Based on AIMD results, we investigated the thermoelectric properties of γ-CuI and monolayer β-CuI from 200 to 500 K. The relaxation times of 2D monolayer β-CuI and

3D γ-CuI were obtained using deformation potential (DP) theory,[60, 61] which is extensively utilized for the analysis of carrier mobility ($\mu$) and relaxation time ($\tau$) in both 2D and 3D materials with the following expressions:

$$\tau_{2D} = \frac{\mu m^*}{e} = \frac{2\hbar^3 C}{3k_B T m^* E_1^2} \tag{1}$$

$$\tau_{3D} = \frac{\mu m^*}{e} = \frac{2\sqrt{2\pi}\hbar^4 C}{3E_1^2 (k_B T m^*)^{3/2}} \tag{2}$$

where $\mu$, $\hbar$, $T$, $m^*$, and $k_B$ are the mobility, reduced Planck constant, temperature, the effective mass, and Boltzmann constant, respectively. $C$ represents the elastic constant of a uniformly deformed crystal, which simulates the lattice distortion activated by the strain. $E_1$ represents the DP constant that indicates the shift of band edges (the conduction band minimum for electrons and the valence band maximum for holes) caused by strain, with considerations for core level correction or vacuum energy level correction in the calculations. For different directions of monolayer β-CuI and γ-CuI, all the calculated parameters are shown in Table 1.

**Table 1** DP Constant $E_1$, elastic constant $C$, effective mass $m^*$, and relaxation time $\tau$ for the monolayer β-CuI and γ-CuI. The units for the elastic constants are $J/m^2$ for 2D crystals and GPa for 3D crystals.

|  | Carrier type | $E_1$ (eV) | $C$ | $m^*$ ($m_e$) | τ (fs) 200 K | 300 K | 400 K | 500 K |
|---|---|---|---|---|---|---|---|---|
| β-CuI, ZZ | electron | -1.51 | 36.10 | 0.32 | 604.78 | 403.19 | 302.39 | 241.91 |
|  | Hole | -2.69 | 36.10 | 0.35 | 171.62 | 114.41 | 85.81 | 68.65 |
| β-CuI, AC | electron | -3.24 | 36.72 | 0.35 | 119.65 | 79.77 | 59.83 | 47.86 |
|  | Hole | -3.66 | 36.72 | 0.49 | 67.33 | 44.89 | 33.66 | 26.93 |
| γ-CuI, a b c | electron | -6.33 | 50.32 | 0.19 | 952.68 | 518.58 | 336.82 | 241.01 |
|  | Hole | -4.70 | 50.32 | 0.41 | 564.07 | 307.04 | 199.43 | 142.70 |

For monolayer β-CuI, the change in conductivity with carrier concentration is determined using relaxation time calculations and $\sigma/\tau$ obtained from BoltzTraP2. The n-type carrier concentration ranges from $1 \times 10^8$ to $2.2 \times 10^{13}$ $cm^{-2}$, and the p-type carrier concentration ranges from $1 \times 10^8$ to $1 \times 10^{14}$ cm$^{-2}$, as shown in Fig. 3(a). For both n-type and p-type doping, the electrical conductivity ($\sigma$) increases with increasing carrier concentration. However, as temperature rises, the increased electron scattering causes the electrical conductivity ($\sigma$) to decrease.[53, 62] At a constant carrier concentration, the electrical conductivity of n-type doped materials exceeds that of p-type doped materials. Additionally, the electrical conductivity in the ZZ direction of monolayer β-CuI is significantly higher than that in the AC direction. Since the $\sigma/\tau$ is calculated using BoltzTrap2, which is almost the same in ZZ and AC directions (see Figure S3), the difference in conductivity in different directions is mainly due to the difference in relaxation time of carriers. According to equation (1) and (2), the carrier relaxation time is determined by $C$ and $E_1$ from DP theory calculations, as well as by the effective mass ($m^*$) and temperature ($T$) (see Supplementary information). For monolayer β-CuI, the $C$ values for the AC and ZZ directions are nearly identical, and the effective mass differences are minimal; the primary factor affecting the relaxation time is the value of $E_1$, as shown in Table 1 and Fig. 3(a).

Fig. 3(b) depict the calculated Seebeck coefficient ($S$) as a function of carrier concentration for monolayer β-CuI at various temperatures. At a specific carrier concentration, the absolute value of the Seebeck coefficient ($|S|$) in the p-type monolayer β-CuI exceeds that of the n-type monolayer β-CuI. The p-type materials typically exhibit a larger Seebeck coefficient ($|S|$) compared to n-type materials, due to the flatter electronic bands near the valence band maximum (VBM).[37] Additionally, in both n-type and p-type systems, the $|S|$ values are nearly identical for both the ZZ and AC directions, and they decrease with increasing carrier concentration. As with most thermoelectric materials, the Seebeck coefficient $|S|$ increases with rising temperature.[63] The power factor ($PF = S^2\sigma$) is a critical parameter for assessing electronic transport properties, which is influenced by both the Seebeck coefficient ($S$) and electrical conductivity ($\sigma$). The $PF$ exhibits a noticeable directional difference: at the same temperature, the maximum $PF$ values for both n-type and p-type monolayer β-CuI are higher along the ZZ direction compared to the AC direction. Specifically, in n-type monolayer β-CuI, the maximum PF along the ZZ direction is approximately four times greater than that along the AC direction, while in p-type monolayer β-CuI, it is about two times greater. At 300 K, the maximum PF values for n-type doped monolayer β-CuI are 13.77 mW m$^{-1}$ K$^{-2}$ in the ZZ direction and 2.87 mW m$^{-1}$ K$^{-2}$ in the AC direction. In contrast, the corresponding maximum PF values for p-type doped monolayer β-CuI are 8.04 mW m$^{-1}$ K$^{-2}$ (ZZ) and 3.26 mW m$^{-1}$ K$^{-2}$ (AC), respectively. It is worth noting that the optimal PF value is comparable to those of 2D thermoelectric materials such as Bi$_2$O$_2$Se (~ 9 mW m$^{-1}$ K$^{-2}$)[75] and ZrSn$_2$N$_4$ (~ 4.5 mW m$^{-1}$ K$^{-2}$)[53].

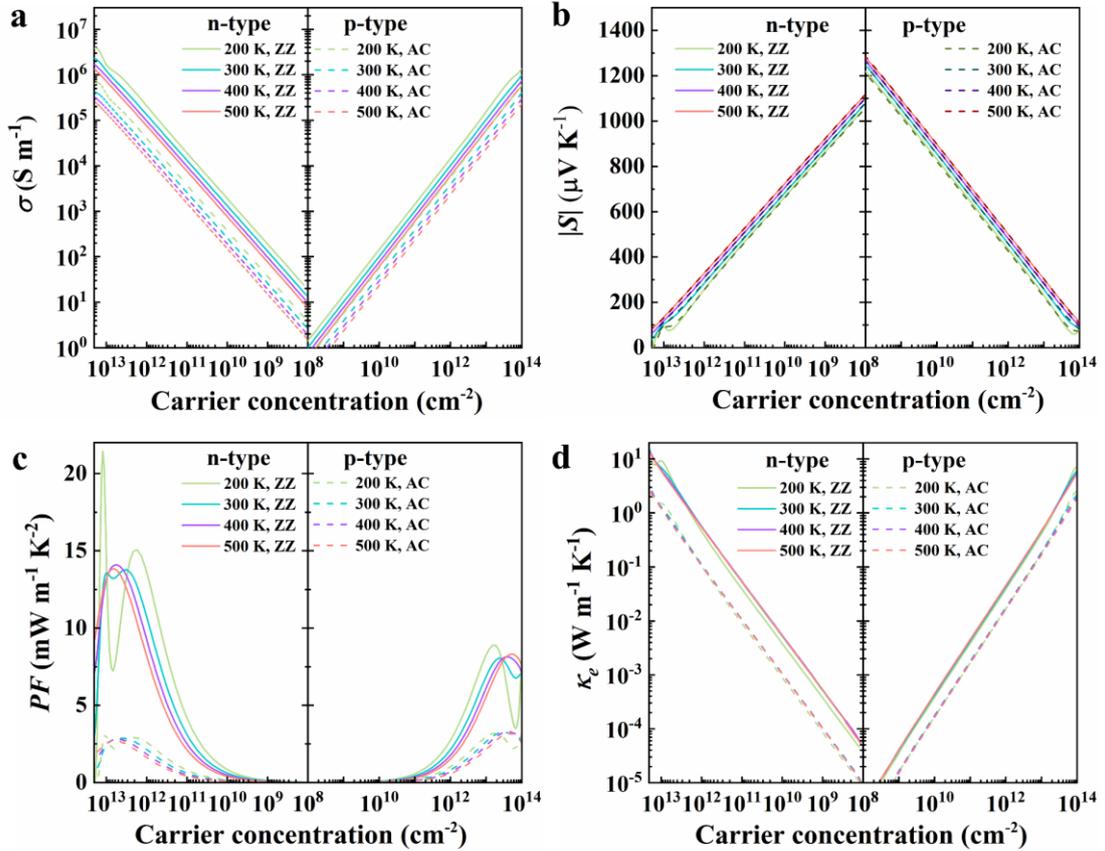

**Fig. 3** The electronic transport coefficients of monolayer β-CuI. (a) The variation of electrical conductivity σ, (b) absolute value of Seebeck coefficient |S|, (c) power factor PF, and (d) electronic thermal conductivity $κ_e$, along the ZZ and AC directions as a function of carrier concentration for n-type and p-type doped monolayer β-CuI at different temperatures.

Furthermore, electronic thermal conductivity is defined as the heat current per unit temperature gradient under open-circuit conditions. However, the electronic thermal conductivity ($k_0/τ$) obtained directly from the BoltzTraP2 code, where ($k_0$) is calculated under closed-circuit conditions. Based on ($k_0/\tau$) and relaxation time ($\tau$), the electronic thermal conductivity ($k_e$) can be corrected using the following equation:[65]

$$k_e = k_0 - T\sigma S^2 \qquad (3)$$

where, $T$ is the temperature, $S$ is the Seebeck coefficient and $\sigma$ is the electrical conductivity. The electronic thermal conductivity, as shown in the Fig. 3(d), shows little variation within the temperature range we studied. However, it exhibits anisotropy, with lower electronic thermal conductivity in the AC direction compared to the ZZ direction, regardless of whether the system is n-type or p-type doped.

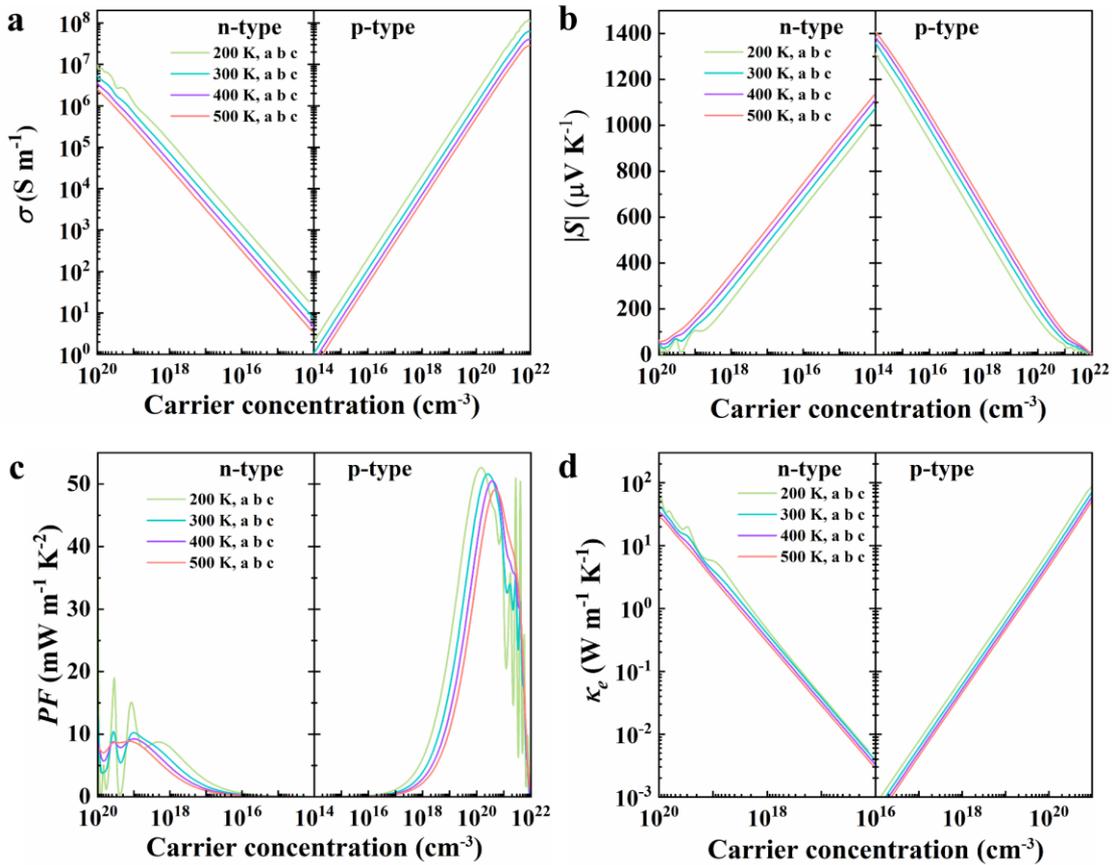

**Fig. 4** The electronic transport coefficients of γ-CuI. (a) The variation of electrical conductivity $\sigma$, (b) absolute value of Seebeck coefficient $|S|$, (c) power factor $PF$, and (d) electronic thermal conductivity $\kappa_e$, along the a, b, and c axes as a function of carrier concentration for n-type and p-type doped γ-CuI at different temperatures.

To the best of our knowledge, no theoretical studies of γ-CuI thermoelectricity have been reported in the literature. We have carried out the calculation of electronic transport coefficients, and the results are shown in Fig. 4. Firstly, in our study, the conductivity, Seebeck coefficient, power factor, and electronic thermal conductivity of the orthorhombic γ-CuI exhibit no directional dependence. It is observed that the maximum power factor *PF* for p-type doping of γ-CuI is higher than that for n-type doping, which contrasts with the behavior observed in monolayer β-CuI. In addition, for γ-CuI, the electronic thermal conductivity decreases with increasing temperature. The optimal *zT* values for γ-CuI and monolayer β-CuI correspond to the electrical conductivity, Seebeck coefficient, power factor, and electronic thermal conductivity at 300 K, as shown in Table S2. It is evident that both the electronic and lattice thermal conductivities of γ-CuI are an order of magnitude larger than those of monolayer β-CuI.

**Thermal transport property**

According to the equation ($zT = S^2 \sigma T / (\kappa_L + \kappa_e)$), thermal conductivity is a crucial factor affecting a material's thermoelectric performance. Materials with excellent thermoelectric properties typically have lower thermal conductivity. Thermal conductivity consists of two parts: electronic thermal conductivity ($\kappa_e$) and lattice thermal conductivity ($\kappa_L$). Previously, we obtained the electronic thermal conductivity for γ-CuI and β-CuI by the Boltzmann transport equation and relaxation time. We assessed the temperature-dependent lattice thermal conductivity of the materials using Phono3py, considering the thickness of the two-dimensional materials, with the results for γ-CuI and monolayer β-CuI shown in Fig. 5(a). As the temperature increases, the phonon mean free path significantly decreases, resulting in a reduction in lattice thermal conductivity.[27] The thermal conductivity of γ-CuI is predicted to be 7.203 W m$^{-1}$ K$^{-1}$, with no difference along the a, b, and c directions of the orthorhombic conventional cell. In contrast, monolayer β-CuI exhibits a lower lattice thermal conductivity, with minor differences between the ZZ and AC directions, at 0.306 and 0.315 W m$^{-1}$ K$^{-1}$, respectively. Xu et al. also reported the thermal conductance of monolayer β-CuI (0.116 W m$^{-1}$ K$^{-1}$ at 300 K) and γ-CuI (0.997 W m$^{-1}$ K$^{-1}$ at 300 K), and carried out detailed mechanism analysis.[27] Mohebpour et al. also reported that monolayer β-CuI's predicted lattice thermal conductivity is approximately 3.75 W m$^{-1}$ K$^{-1}$.[30] Our calculated structure is basically consistent with the reported results: monolayer β-CuI has ultra-low lattice thermal conductivity, which is significantly lower than γ-CuI. This also indicates that monolayer β-CuI has lower thermal conductivity and may have better thermoelectric properties than γ-CuI. To gain a deeper understanding of the mechanisms underlying the low lattice thermal conductivity of monolayer β-CuI and the significant differences in lattice thermal conductivity compared to γ-CuI, we compared their phonon group velocities and lifetimes. As illustrated in Fig. 5(b) and (d), both the phonon group velocity and relaxation time are small for monolayer β-CuI, which is consistent with the findings of Xu et al. and Mohebpour et al.[27, 30] Moreover, Xu et al. provided a deeper mechanistic explanation, noting that the total absolute value of Grüneisen parameters for monolayer β-CuI (1.6055) is significantly higher than that for γ-CuI (0.4828), and that the phonon mean free path of monolayer CuI is shorter compared to γ-CuI.[27] It is worth noting that we also consider the electron thermal conductivity and provide a comprehensive assessment of the overall thermal conductivity, which reveals that monolayer β-CuI exhibits extremely low thermal conductivity.

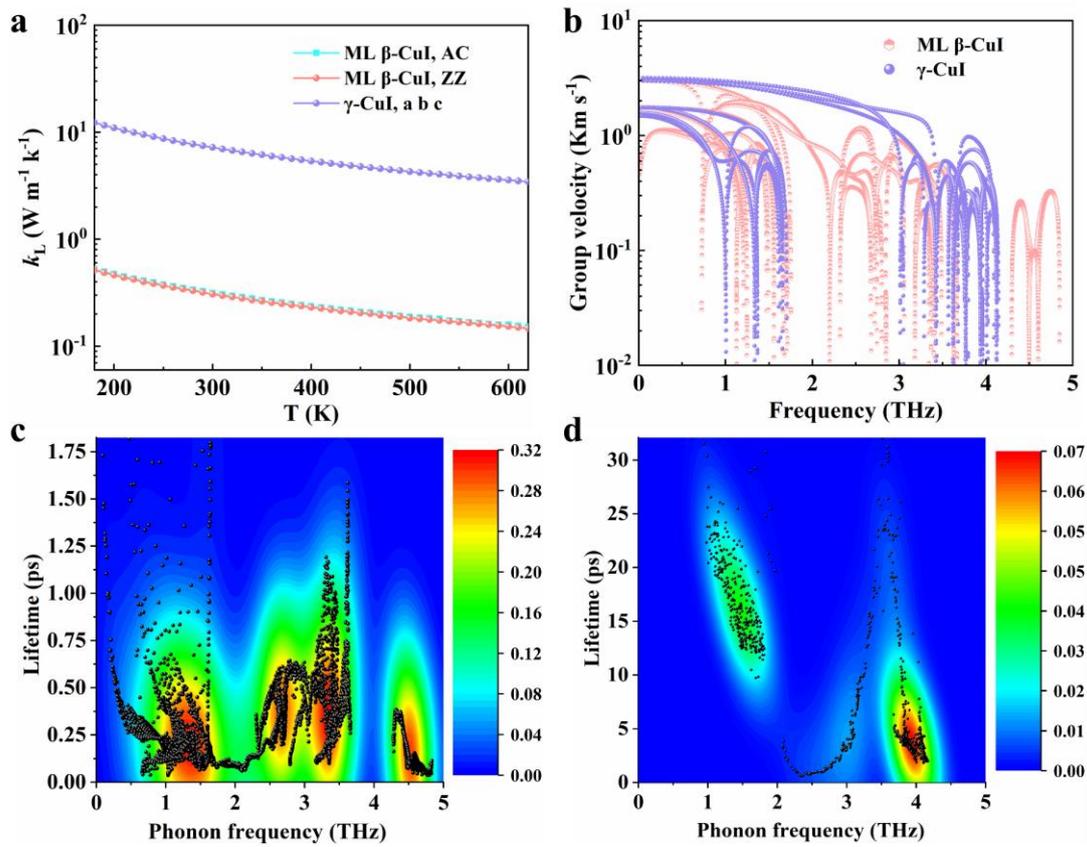

**Fig. 5** (a) Temperature-dependent lattice thermal conductivity of monolayer β-CuI and γ-CuI. (b) Group velocities of the phonons as a function of frequency. Phonon relaxation time as a function of frequency for (c) monolayer β-CuI and (d) γ-CuI. The color bar represents the density of data points.

**Figure of merit**

The conversion efficiency of thermoelectric materials is quantitatively described by the dimensionless parameter $zT$. Materials with a large $zT$ value are recognized as efficient thermoelectric materials. The variation of $zT$ values with carrier concentration along different directions and at various temperatures for 2D monolayer β-CuI and 3D γ-CuI is illustrated in Figs. 6(a) and 6(b). The value of $zT$ increases with carrier concentration, peaks, and subsequently decreases with further increases in concentration. In the investigated system, the $zT$ value exhibits a rising trend with temperature and reaches its peak at 500 K. Notably, the $zT$ values of β-CuI show a significant difference between the ZZ and AC directions, with the $zT$ value in the ZZ direction being higher than in the AC direction at the same temperature. At 300 K, the peak $zT$ values for n-type doped monolayer β-CuI are 4.10 in the ZZ direction and 1.52 in the AC direction, with carrier concentrations of $5.18 \times 10^{11}$ cm$^{-2}$ and $1.34 \times 10^{12}$ cm$^{-2}$, respectively. For p-type doped monolayer β-CuI are 2.98 in the ZZ direction and 1.58 in the AC direction, with carrier concentrations of $5.44 \times 10^{12}$ cm$^{-2}$ and $7.93 \times 10^{12}$ cm$^{-2}$, respectively. The monolayer β-CuI exhibits excellent low-temperature thermoelectric characteristics and significant application potential, as evidenced by its room temperature $zT$ value exceeding 1.50.

The maximum $zT$ value of γ-CuI at room temperature (300 K) is 0.30 for n-type doping and 1.09 for p-type doping. Bae et al. fabricated CuI films demonstrating an excellent electrical conductivity of 207.6 S cm$^{-1}$ and a power factor of 673.3 μW m$^{-1}$K$^{-2}$, leading to a high $zT$ value of 0.31.[43] Yang et al. developed a transparent, flexible thermoelectric material based on non-toxic, earth-abundant p-type copper iodide thin films, achieving a substantial thermoelectric figure of merit with a $zT$ value of 0.21 at 300 K.[42] Coroa et al. reported a maximum $zT$ value of 0.29 for single CuI thin films with a thickness of approximately 300 nm.[47] Currently, the experimentally measured $zT$ values for γ-CuI remain lower than the theoretical values, indicating a need for further refinement and exploration. Additionally, at room temperature, β-CuI demonstrates superior *zT* values compared to γ-CuI for both p-type and n-type doping, mainly due to significant differences in thermal conductivity (see Table S2). Most importantly, monolayer β-CuI exhibits average *zT* values of 2.28 for p-type and 2.81 for n-type doping at room temperature. Furthermore, over the temperature range of 200 to 500 K, the optimal *zT* values of β-CuI consistently exceed those of γ-CuI at each temperature, as illustrated in Fig. 6(c) and (d). This also indicates that β-CuI exhibits better low-temperature thermoelectric properties than γ-CuI.

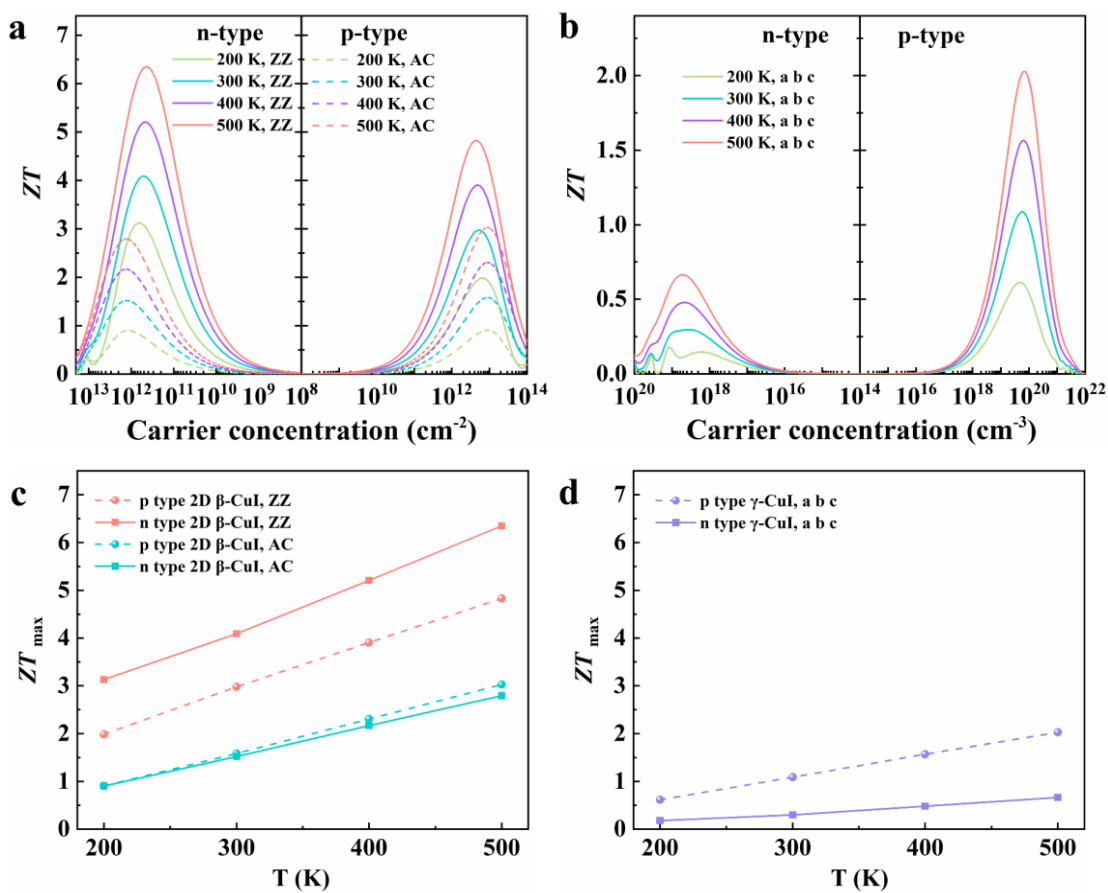

**Fig. 6** The Calculated figure of merit *zT* of (a) β-CuI monolayer and (b) γ-CuI as a function of carrier concentration. The optimal *zT* value of (c) β-CuI monolayer and (d) γ-CuI as a function of temperature.

## CONCLUSIONS

In summary, this study utilizes first-principles calculations, Boltzmann transport theory, and AIMD simulations to investigate the structural, electronic, and thermoelectric transport properties of monolayer β-CuI. Firstly, monolayer β-CuI shows thermodynamic stability near room temperature. At 500 K, it transitions to a superionic state, and at 700 K and 900 K, it enters a liquid phase. Additionally, a comparison of the thermoelectric properties between monolayer β-CuI and γ-CuI reveals that monolayer β-CuI exhibits superior thermoelectric performance at low temperatures (200 – 500 K). The thermoelectric properties of monolayer β-CuI exhibit anisotropy along the ZZ and AC directions. At 300 K, the $zT$ values for p-type and n-type doping are 2.98 and 4.10, respectively, along the ZZ direction, while along the AC direction, they are 1.58 and 1.52. This indicates that the thermoelectric performance along the ZZ direction is superior to that along the AC direction. In addition, compared to monolayer β-CuI, γ-CuI exhibits lower $zT$ values at room temperature, with $zT$ values of 1.09 for p-type doping and 0.30 for n-type doping. For both monolayer β-CuI and bulk γ-CuI, the $zT$ values increase with temperature, reaching their maximum at 500 K. At 500 K, the optimal $zT$ values for monolayer β-CuI are 4.83 (p-type) and 6.35 (n-type) along the ZZ direction, and 3.03 (p-type) and 2.79 (n-type) along the AC direction. For γ-CuI at 500 K, the optimal $zT$ values are 2.03 (p-type) and 0.67 (n-type). At the same temperature, the $zT$ values of monolayer β-CuI are consistently higher than those of γ-CuI, indicating that monolayer β-CuI exhibits superior thermoelectric performance compared to γ-CuI. The findings highlight the significant potential of the β-CuI monolayer as a low-temperature thermoelectric materials that is both cost-effective and environmentally friendly.

## METHODS

**DFT Parameters**

All first principles calculations based on DFT were performed with the Vienna ab initio Simulation Package code, using projector augmented-wave (PAW) pseudopotential method to account for interactions between core and valence electrons.[66, 67] The generalized gradient approximation of Perdew–Burke–Ernzerhof (GGA-PBE) has been selected to handle the exchange–correlation energy.[66] A plane-wave basis set with a kinetic-energy cut-off of 520 eV was used to expand the wave function of valence electrons (5s2 p5 for I, and 3d10 4s1 for Cu). The structural relaxations were performed by computing the Hellmann–Feynman forces within total energy and force convergences of $10^{-7}$ eV and $10^{-5}$ eV Å$^{-1}$, respectively. Gamma-centered Monkhorst–Pack grids of 15 × 15 × 1 and 14 × 14 × 14 were used for monolayer β-CuI and γ-CuI model. A 20 Å vacuum space was set in the normal direction to prevent interaction in adjacent β-CuI monolayers.[32] To accurately represent the electronic structure and transport coefficients, the Heyd-Scuseria-Ernzerhof (HSE06) hybrid functional is employed for describing the exchange-correlation energy.[68] To achieve precise effective mass values, the effects of spin-orbit coupling (SOC) are also factored into consideration. The electronic transport coefficients for both γ-CuI and β-CuI are determined using semiclassical Boltzmann theory within the relaxation-time approximation in the BoltzTraP2 code.[69] A series of test calculations was performed to identify the optimal k-point sampling for evaluating the transport coefficients (see Figure S2, S3 and S4). High k-point meshes are employed: 24 × 14 × 1 for the orthorhombic conventional cell of γ-CuI and 11 × 11 × 11 for the orthorhombic conventional cell of β-CuI. The relaxation time ($\tau$) is computed based on the deformation potential (DP) theory.[60] The Phonopy and Phono3py packages utilize second- and third-order interatomic force constants as inputs for computing phonon dispersion and lattice thermal conductivity.[70, 71] We used the VASPKIT code for post-processing of the VASP calculated data,[72] and the crystal structure was visualized using VESTA software.[73]

**AIMD Simulations**

To evaluate the thermal stability of the monolayer β-CuI, all Ab Initio molecular dynamic (AIMD) simulations were done using the CP2K (version 8.1).[74] The density functional implementation in Quickstep was based on a hybrid Gaussian plane wave scheme. Orbitals were described by an atom centered Gaussian-type basis set, and a plane wave basis set with a cutoff set to 500 Ry was used to re-expand the electron density in the reciprocal space. PBE functional with Grimme's dispersion correction was used. The core electrons were represented by analytic Goedecker–Teter–

Hutter (GTH) pseudopotentials.[75] For valence electrons (3d, 4s for Cu, and 5s, 5p for I), the Gaussian basis sets were double-ζ basis functions with one set of polarization functions (DZVP-MOLOPT-SR-GTH).[76] AIMD simulations were carried out with CP2K simulation package for more than 20 ps at a time step of 1 fs, which was performed on super cell of monolayer β-CuI, as shown in Figure 2. At the start of the AIMD simulations, the monolayer β-CuI samples were assigned an initial temperature of 300 K according to a Boltzmann distribution. The samples were then heated up to the desired temperature by annealing velocities (1.001). For all the AIMD trajectories, the initial 5 ps (5000 steps) was regarded as the equilibration period. Canonical (NVT) ensemble and Nose–Hoover thermostats were set to 300, 500, 700, and 900 K, respectively. Note that due to the large size of the supercells, only $\Gamma$ point was used in all calculations, and an energy convergence of $1 \times 10^{-6}$ eV was chosen.

## DATA AVAILABILITY

The data that support the findings of this study are available from corresponding author upon reasonable request.

## COMPETING INTERESTS

The authors declare no competing interests.

## ACKNOWLEDGEMENTS

We thank Professor Haiping Fang for his constructive suggestions. This work was supported by the Startup Fund of Wenzhou Institute, University of Chinese Academy of Sciences (No. WIUCASQD2021014 and WIUCASQD2023017), the Fundamental Research Funds for the Central Universities, the Scientific Research and Developed Funds of Ningbo University (No. ZX2022000015).

## AUTHOR CONTRIBUTIONS

B.Q.P and L.C. conceived the idea. B.Q.P and Y.S.L. conducted the simulation and analysis. All authors participated in the writing and correction of the manuscript.